\documentstyle[aps,12pt]{revtex}

\begin{document}
\title{Giant Sommerfeld coefficient in the heavy-fermion YbBiPt}
\author{Z.~Ropka}
\address{Center for Solid State Physics, \'{s}w. Filip 5, 31-150 Krak\'{o}w, Poland.\\
email: sfradwan@cyf-kr.edu.pl}
\maketitle

\begin{abstract}
It has been derived that the Sommerfeld coefficient $\gamma $ as large as 25
J/K$^2$mol can be theoretically accounted for provided the charge f-electron
fluctuations are substantially suppressed. This result enables the
understanding of the heavy-fermion YbBiPt ($\gamma $ =8 J/K$^2$mol) as the
spin fluctuator and should stimulate the experimental specific-heat research
at low temperatures.


\end{abstract}

In 1991 Fisk et al. [1] discovered that YbBiPt has the linear-temperature
specific-heat coefficient (Sommerfeld coefficient) $\gamma $ of 8 J/K$^2$%
mol. It is the largest value measured up to now - in the highest
heavy-fermion systems a value up to 1.6 J/K$^2$mol has been found (CeAl$_3$%
). The origin of such the giant value is still the subject of extensive
theoretical studies and long-lasting discussions [1-5]. In fact, this
discussion concentrates on the role played by 4f electrons and the way how
to treat them. One school considers the f electrons as the part of the core
whereas the second one treats them as itinerant band--like electrons.

The aim of this Letter is to show that a giant value of 25 J/K$^2$mol for
the Sommerfeld coefficient can be derived using the method described in Ref.
2 provided the charge f-electron fluctuations are strongly suppressed.

In the theoretical paper of Ref. 2 it has been concluded that the
band-structure LSDA+U calculations provide the single-particle density of
states that i) is peak-like at the Fermi level E$_F$ and ii) its value of $%
\mathop{\rm about}
$ 200 states per eV yields the Sommerfeld coefficient $\gamma $ of 0.25 J/K$%
^2$mol. Moreover, in order to make this value closer to the experimental one
of 8 J/K$^2$mol in Ref. 2 iii) an enhancement by the factor (1-n$_f$)$^{-1}$
has been invoked. This factor has been introduced to the charge-fluctuation
model for heavy-fermion phenomena in Ref. 4 and is due to the suppression of
the f occupation fluctuations. In Ref. 2 the n$_f$ has been assumed as 0.9.
It increases the bare LSDA+U Sommerfeld-coefficient value by factor of 10 to
2.5 J/K$^2$mol. The large density of states at E$_F$ in the heavy-fermion
system YbBiPt found in Ref. 1 has been recently confirmed by the ab initio
self-interaction correction to the local-spin density approximation
(SIC-LSDA) calculations of Temmerman et al. [7], though in this
approximation the f electrons are not part of the core.

According to me in the approach described in Ref. 2 there is no
justification for the assumed value of 0.9 for n$_f$. In fact, basing on the
original approach to YbBiPt [1] the value of n$_f$ of 0.99 seems to be much
more realistic. It is also in agreement with the suppresion of the charge
fluctuations already in temperatures above 10 K where the full trivalent
ytterbium moment is seen in the susceptibility vs temperature plot [1,3].
Such the value for n$_f$ yields so giant Sommerfeld coefficient as 25 J/K$^2$%
mol. I predict that such the large Sommerfeld-coefficient value will be
experimentally detected for ytterbium/cerium/samarium systems with the
characteristic temperatures below 0.1 K. (In YbBiPt the Sommerfeld
coefficient of 8 J/K$^2$mol occurs below 0.4 K). Values of such order has
been found already for $^3$He below 0.2 K [8,9].

In conclusion, I have derived that the Sommerfeld coefficient as large as 25
J/K$^2$mol can be theoretically accounted for provided the charge f-electron
fluctuations are substantially suppressed. This result enables the
understanding of the heavy-fermion YbBiPt ($\gamma $ =8 J/K$^2$mol) as
almost spin-fluctuating system (the f electrons are part of the core) and
should stimulate the experimental specific-heat research at low
temperatures. According to Refs 5 and 6 these spin fluctuations are related
with the atomic scale Kramers doublet ground state of the trivalent
ytterbium ions.

\end{document}